\documentclass[11pt,twoside]{article}

\usepackage{asp2006}
\usepackage{epsf}

\begin{document}

\title{The Morphological Type Dependence of K-band Luminosity Functions}   %%% Fill in title

\author{Nick Devereux \& Paul Hriljac}
\affil{Embry-Riddle Aeronautical University, Prescott, AZ}

\author{S.P. Willner \& M.L.N. Ashby}
\affil{Harvard-Smithsonian Center for Astrophysics}

\author{C.N.A. Willmer}
\affil{University of Arizona}

\begin{abstract} %%% Abstract to run on from here.

Differential 2.2${\micron}$ (K-band) luminosity functions are presented for a complete sample of 1570 nearby (Vgsr ${\le}$ 3000 km/s, where Vgsr is the velocity measured 
with respect to the Galactic standard of rest), bright (K ${\le}$ 10 mag), galaxies segregated by visible morphology. 
The K-band luminosity function for late-type spirals follows a power law that rises towards low luminosities whereas the K-band
luminosity functions for ellipticals, lenticulars and bulge-dominated spirals are peaked
with a fall off at both high and low luminosities.
However, each morphological type (E, S0, S0/a-Sab, Sb-Sbc, Sc-Scd) contributes approximately equally to the overall K-band luminosity density in the local universe, and by inference, the stellar mass density as well.

\end{abstract}

%%% MAIN BODY OF TEXT GOES HERE. CONSULT "INSTRUCTIONS FOR AUTHORS USING
%%% LATEX2E MARKUP", SECTIONS 2.3-2.6 FOR HELP WITH EQUATIONS, FIGURES,
%%% AND TABLES.

%\section{}   %%% Top level section head (remove "%" symbol)
%\subsection{}   %%% Second level section head (remove "%" symbol)
%\subsubsection{}   %%% Lowest level section head (remove "%" symbol)
%\section*{}    %%% Unnumbered top level section head (remove "%" symbol)
%\subsection*{}   %%% Unnumbered second level section head (remove "%" symbol)

\section{Introduction}

The Two Micron All Sky Survey \citep[2MASS,][]{Skr06} constitutes a unique resource that has been exploited in recent years to produce near-infrared luminosity functions for galaxies with ever greater precision \citep{Col01, Koc01, Bel03, Eke05, Jon06}. The K-band luminosity function provides a key constraint in understanding galaxy evolution in the context of Lambda Cold Dark Matter (${\Lambda}$CDM) cosmology by virtue of the fact that the zero redshift K-band luminosity  function traces the stellar mass accumulated in galaxies at a wavelength where interstellar extinction is minimal \citep{Dev87, Bad01, Bel03}.  

The current paradigm, constrained, in part by the K-band luminosity function \citep[e.g.,][]{Ben03}
has galaxy disks forming through a combination of cold gas accretion and feedback \citep[e.g.,][]{Dut08} with bulges resulting from mergers \citep[e.g.,][]{Mas08}. Thus, the diversity of visible morphologies seen today among galaxies represents the culmination of multiple evolutionary paths. These end points in galaxy evolution are captured in a taxonomy devised by \citet{Hub36} and refined by \citet{deV59}, that is based on the relative prominence of the stellar bulge and the degree of resolution of the spiral arms. 

Nearby galaxies were identified using HYPERLEDA; a web-based interface (http://leda.univ-lyon1.fr) that provides access to the Principal Galaxy Catalog \citep{Pat03}. The 2MASS counterparts were identified on the basis of positional coincidence with the Extended Source Catalog \citep[XSC,][]{Jar00}
A volume-limited sample was defined for further study, hereafter the K10/3000 sample, comprising 1604 galaxies with K  ${\le}$ 10 mag, Vgsr ${\le}$ 3000 km/s,  and ${|b|}$  ${>}$ 10 degrees. The adopted K-band magnitudes are those measured within the 20 mag/(arc sec)$^{2}$ elliptical isophote; the parameter k${_{-}}$m${_{-}}$k20fe in the 2MASS XSC.  

The principal aim of this project, described in more detail in \cite{Dev09}, is to use nearby galaxies to define the first benchmark K-band luminosity functions for galaxies segregated by visible morphology.
The luminosity function calculation employs the non-parametric maximum likelihood method
of \cite{Cho86}.
Our study improves
on \citet{Mar94,Mar98,Bin88,Efs88,Koc01} and \citet{Bel03} by limiting the sample to include only
nearby galaxies, which have the most reliable morphological assignments, and by
using the most recent galaxy distance
determinations in conjunction with near infrared K-band magnitudes that correlate with stellar mass.

\begin{figure}[!ht]
\plotone{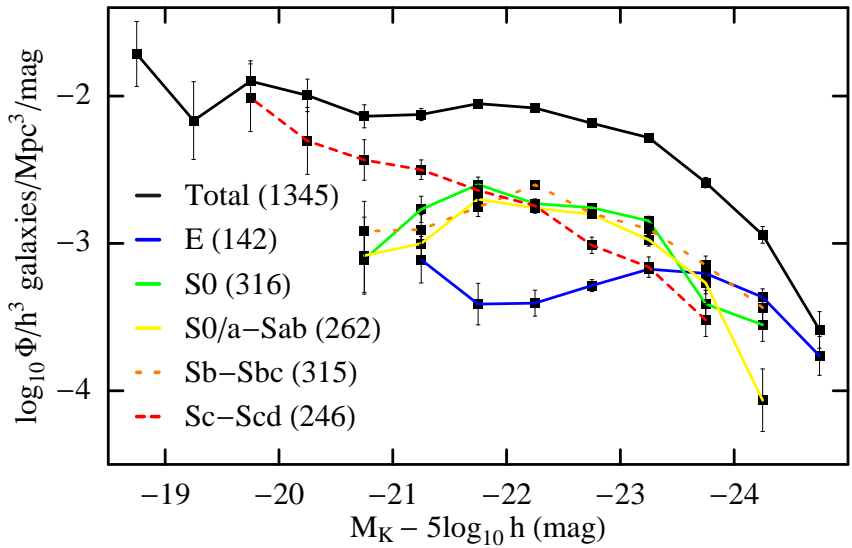}
\caption{{\bf K-band isophotal luminosity functions for 1345 galaxies in the K10/3000 sample segregated by visible morphology. The error bars reflect statistical uncertainties only.  
The binning procedure inherent to the Choloniewski method excludes some galaxies which is why the total number of galaxies is not the sum of the number of galaxies within each Hubble type, and why the total number of galaxies in the plot is less than the total in the sample.}}
\end{figure}

\section{Morphological Type Dependence of the K-band Luminosity Function}

Figure 1 shows that galaxies of different morphological type have different luminosity functions and no
type mimics the shape of the total luminosity function.
Ellipticals dominate the space density at high luminosities, whereas late-type (Sc - Scd) spirals dominate the space
density at low luminosities. Lying between these two extremes are the lenticular galaxies and the bulge-dominated spirals (S0/a - Sbc). 

The {\it total} K-band luminosity density, calculated by integrating the total luminosity function yields (5.8 ${\pm}$ 1.2) ${\times}$ 10${^8}$  ${\it h}$ L${_\odot}$ Mpc$^{-3}$. Elliptical galaxies contribute ${\sim}$ 16 ${\pm}$ 3${\%}$ of the total. Lenticulars and bulge-dominated spirals combined  contribute ${\sim}$ 68 ${\pm}$ 14${\%}$ of the total, or ${\sim}$ 22 ${\pm}$ 4${\%}$ for each sub-group (S0, S0/a-Sab, Sb-Sbc). Finally, the late-type spirals  contribute  ${\sim}$ 16 ${\pm}$ 3${\%}$ of the total. {\it Overall, to a good approximation, one could say that each Hubble type (E, S0, S0/a-Sab, Sb-Sbc, Sc-Scd) contributes equally to the overall K-band luminosity density in the local universe.} Using information provided
in \cite{Bel03}, one can predict that the M/L ratio measured in the K-band will not vary systematically by more than ${\sim}$ 7\% between E and Scd galaxies. Consequently, each morphological type contributes approximately equally to the stellar mass density as well. 

As far as the shape of the luminosity functions, late-type spirals follow a power law that rises towards low luminosities, whereas the ellipticals, lenticulars and bulge-dominated spirals (S0/a - Sbc)  
are peaked with a fall off at both high {\it and} low luminosities. 
Our results concerning the morphological type dependence of K-band luminosity functions differ from previous studies. \cite{Koc01} and \cite{Bel03} divided their sample into just two broad categories; {\it early} and {\it late}, based on visual classifications and the SDSS light concentration index. They found little difference in form between the luminosity functions for the different types, which reflect that of the total luminosity function. Our results, based on a more comprehensive segregation according to the visual morphological classification scheme of \cite{deV59}, reveal
significant differences between the luminosity functions for the different types, none of which mimic the shape of the total luminosity function.

\section{Discussion}

Our principal new result is that {\it the shape of the K-band luminosity functions depend significantly on galaxy visible morphology}. It may be more than a coincidence that the functional forms distinguish between {\it bulge} dominated and  {\it disk} 
dominated systems. Evidently, there are at least two quite distinct galaxy formation mechanisms at work to produce the diversity of morphological types seen in the local universe. The next step, of course, is to establish what the formation mechanisms are exactly, which will require modeling the {\it differential} luminosity functions in the context of hierarchical clustering scenarios \citep[e.g.,][]{Col00,Ben03}. 

Semi-analytic models have already revealed that a combination
of cold gas accretion \citep{Wei04} and feedback \citep{Opp06} can flatten the slope of the halo mass function to match that seen for galaxy disks. 
Such models are also able to reproduce the peaked luminosity functions observed for ellipticals and bulge-dominated spirals by incorporating major mergers \citep{Bar92,Hop08}. Thus, there is promise
that the morphological dichotomy revealed by the K-band luminosity functions may be understood within the context of ${\Lambda}$CDM cosmology (Benson 2008, private communication).
On the other hand, dwarf ellipticals pose a problem; even though they may be structurally related to their more luminous counterparts \citep{Kor08},  our results show that they have a distinct luminosity function as noted previously by \cite{Bin88}.

%%\acknowledgements %%% Text of acknowledgements runs on after this command.

%%% THE BIBLIOGRAPHY
%%%
%%% CONSULT SECTION 3 OF "INSTRUCTIONS FOR AUTHORS" FOR HOW TO USE NATBIB.
%%% AUTHORS ARE ENCOURAGED TO USE EITHER THE "THEBIBLIOGRAPY" ENVIRONMENT
%%% BY UNCOMMENTING (DELETING THE "%" SYMBOL) THE COMMANDS BELOW, OR BY
%%% USING THE BIBTEX ENVIRONMENT. TO FIND OUT WHICH IS APPLICABLE TO YOUR
%%% CONTRIBUTION, CONSULT THE VOLUME EDITORS FOR YOUR PROCEEDINGS.
%%%

\end{document}